# Development of an Interface for Using EGS4 Physics Processes in Geant4

K.Murakami, K.Amako, H.Hirayama, Y.Namito, T.Sasaki
*KEK, Tsukuba, Ibaraki 305-0801, Japan*

M.Asai, T.Koi
*SLAC, Stanford, CA 94025, USA*

As simulation system, the variety of physics processes implemented is one of the most important functionalities. In that sense, Geant4 is one of the most powerful simulation toolkits. Its flexibility and expansibility brought by object-oriented approach make it possible for us to easily assimilate external simulation packages into the Geant4 system as modules of physics processes. We developed an interface for using EGS4, which is another of the most well-known simulation package for electromagnetic physics, in Geant4. By means of this interface, EGS4 users can share Geant4 powerful resources, such as geometry, tracking, etc. It is also important that it can provide a common environment for comparison tests between EGS4 and Geant4. In this paper, we describe our design and implementation of the interface.

## 1. INTRODUCTION

Geant4 is a toolkit for detector simulation of the passage of particles through matter. It is composed of geometry, material, tracking particles, particle interaction, detector response, event, run, visualization and user interfaces [1]. Exploiting an object-oriented methodology and C++ programming language, Geant4 enjoys rich flexibility and expansibility as generic simulation system. Particularly in terms of physics process, Geant4 gives clear and definite protocols between particle and process, and tracking and process, so that users can make a choice of models for each process or easily implement additional users' own processes. On the other hand, EGS4 is another well-established simulation code [2]. Physics processes implemented in EGS4 are confined to electromagnetic physics of electron and photon. But it is widely used as a standard especially in medical research area.

In general, comparing between different simulation tools often raises controversial issues in geometry description, cut values and so on. In order to minimize these kinds of problems, our attention is focused on interfacing these well-established and commonly-used simulation codes and making them co-work in a single framework. Practically, taking advantage of the capability of Geant4 as a framework, we have developed a plug-in interface for using EGS4 as a module of Geant4 physics process. By means of this interface, it is expected that EGS4 users can share Geant4 powerful resources, such as geometry description, tracking, etc. And also it can provide a common testing environment for comparison tests in users' realistic application level.

In the following, we describe our design and implementation of the interface, and the results of basic benchmarks are also presented.

## 2. SYSTEM ANALYSIS AND DESIGN

First of all, our design policy is set. We should care for not to change algorithms and prescriptions which are applied in the EGS4 system in order to get the same output as obtained by using EGS4 itself. According to our policy, we make careful system analysis for each component of EGS4 and fix whether it can be replaced with a corresponding Geant4 component or should be left as it is. Analysis of each functional component is described in the following subsections, and summarized in Table 1.

Table 1: Summary of which package is used in each functional component

| Component | Which package? |
|---|---|
| Control Framework | Geant4 |
| Geometry and Tracking | Geant4 |
| Material | Geant4/EGS4 |
| Cross Section Table | EGS4 (PEGS) |
| Physics Process | EGS4 |
| Cutoff | EGS4 |

### 2.1. Geometry and Tracking

Geometry description model is very different from each other. In Geant4, describing geometry is completely separated from the tracking component in users' implementation level. Particle transportation is carried out internally in a generic way[1]. Geometry model is based on the mother-daughter concept. Each volume is placed in its mother volume hierarchically, so that users can implement their own detector geometry intuitively.

In EGS4, on the other hand, geometry description is closely coupled with users' implementation of particle transportation. To implement geometry, users have to calculate the distance between the current location and the closest surface and transport a particle in their own codes. In other words, EGS4 provides no generic way of describing geometry and tracking particles. Therefore we do not have any difficulties in making use of Geant4 functionalities for describing geometry and tracking particles.

---

[1] Practically, particle transportation is implemented as a kind of process.

## 2.2. Process and Tracking

The next consideration is focused on how a physics process is invoked in tracking, that is, a way of step size calculation and process selection. In Geant4, each process has its own fate in terms of number of interaction length. Then, a process having minimum fate will be invoked. In EGS4, the next step size is decided based on the total cross section. Then, which process will be applied is chosen at the rate of branching fraction of a process. In principle, these two treatments function in the same manner. The remaining small difference is fixed in implementation level. Practically, we implement EGS4 physics processes as single Geant4 processes for electron and photon respectively.

Subroutines of physics processes in EGS4 are written in FORTRAN (Mortran in practice). Each of them is reused as it is. This is because the already existing FORTRAN codes are working well and it is more convenient to keep independence from each package for future updates.

## 2.3. Material and Cross Section Table

In the EGS4 system, defining materials and calculation of cross section tables have to be prepared by PEGS, which is an external program, prior to execution of users' programs. An output file of PEGS is read in an EGS4 program at initialization phase. We make a design decision to share this procedure for defining materials and calculating cross section tables.

Information described as material attributes is slightly different. In Geant4, in most cases, material information has only properties of material itself, such as atomic number, composition, density, etc. In EGS4, additional information associated with process such as cutoff energy, parameters and coefficients of formulas, option flags specific to processes are also given. Therefore, our interface is supposed to manage materials in terms of both Geant4 and EGS4. Besides Geant4 material classes, classes describing materials defined in EGS4 are designed. In addition, a mapping table between Geant4 materials and EGS4 materials is in need to convert a Geant4 material to a corresponding EGS4 material, and vice versa.

## 2.4. Cutoff

The word "cutoff" has two meanings. One is "tacking cut"; particles below the cutoff energy will be discarded. The second meaning is "production cut"; only particles above the cutoff will be generated as secondary particles. In Geant4, only production cut is applied, and particles basically will be transported until they have zero kinetic energy. In EGS4, on the contrary, both cutoffs are applied.

A way of setting cutoff values is also different. In Geant4, only one value is set in terms of range, and it is internally converted to energy for each combination of material and particle species. On the other hand, in EGS4, a cutoff parameter is specified for each material. So, we manage cutoff values in material description.

According to our policy, our implementation should be taken care so as not to change the cutoff treatment carried out in EGS4. Therefore, we bypass cutoff treatment in Geant4 and set and apply cutoff in the EGS4 style.

## 3. IMPLEMENTATION

### 3.1. Material

There are two additional classes presented to describe materials used in EGS4; *EGS2G4PegsMaterial* and *EGS2G4Material*. *EGS2G4PegsMaterial* describes a material defined in a PEGS output file. Its attributes consist of name, index of a material array in the EGS4 common block, density, energy range of cross section tables (*AP*, *UP*, *AE*, *UE*) and some user flags. Its objects will be instantiated from a PEGS output at initialization time. *EGS2G4Matetial*, which is referred by EGS4 processes, represents a practical medium assigned to a volume. It has user parameters, like cutoff (aka *ECUT*, *PCUT*), density, process switch, etc in addition to a reference to an object of *EGS2G4PegsMaterial*. A correspondence between *EGS2G4Material* and *G4Material* has to be given as a form of a mapping table (*g4-egs4 material map*) at initialization. This table is used at tracking time to retrieve an *EGS2G4Material* object corresponding to a *G4Material* one of a current volume.

### 3.2. Physics Process

Geant4 is designed to allow users to implement new processes in a generic way. All actual process classes inherit from *G4VProcess*, in which definite protocols to invoke interactions from the tracking side are defined. All processes are classified into the following process types; "AtRest", "AlongStep", "PostStep" and any combinations of these types. Each type of process has two mandatory methods to be implemented. *GetPhysicalInteractionLength(GPIL)()* limits the step size and calculates when an interaction will occur. *DoIt()* gives actual changes of a particle and creates secondary particles.

In EGS4, tracking electrons and photons is handled in the FORTRAN subroutines, *ELECTR* and *PHOTO* respectively. Each actual physics process is called in these subroutines at the rate of its cross section.

Practical implementation is carried out according to the Geant4 scheme. As previously mentioned, single Geant4 processes per electron and photon are implemented; *EGS2G4eProcess* for electron and *EGS2G4gProcess* for photon, inheriting from *G4VProcess*. They are implemented as a continuous-discrete-at-rest process and a discrete process respectively. The corresponding EGS4 subroutines (*ELECTR* and *PHOTO*) are divided into two parts so as to match with the protocols of Geant4 process. One is a part of calculating *GPIL()*, and the other is to describe actual reaction of processes, i.e. *DoIt()*. The divided parts are written in Mortran codes and wrapped with C++ because it is much easier and safer to handle the original codes.

**THMT006**

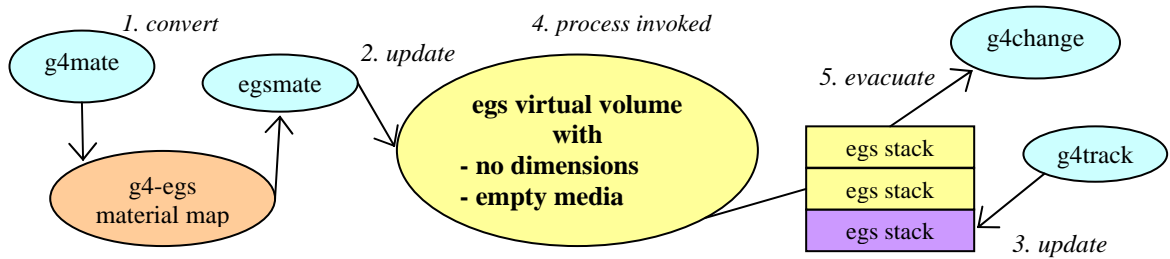

Figure 1: Schematic view of the work flow at tracking time

The list of processes currently covered is summarized in Table 2. At the first target, we start with connecting a plain version of EGS4 package. Some extension components like PRESTA are not included yet. It is noted that the current implementation of multiple scattering is described only by flipping momentum direction. In order to include lateral displacement, which is included in the PRESTA extension, we have to set a safety limit on lateral displacement not to push an electron out to the next volume. This is the next target of our plan.

Table 2: Process list currently covered

| Electron processes | Photon processes |
|---|---|
| multiple scattering | Rayleigh scattering |
| continuous energy loss | pair production |
| Bremsstrahlung | photo-electric effect with subsequent fluorescence |
| Møller scattering | Compton scattering |
| positron annihilation (in-flight / at-rest) | |

With regard to cutoff treatment, a particle below cutoff energy will be stopped and killed with depositing its kinetic energy. In the case of positron, a positron will be stopped and followed by annihilation to two photons.

### 3.3. How It Works

In this subsection, how the interface works is described from the tracking view point. We let particles travel in parallel worlds; Geant4 and EGS4 geometries. A Geant4 world is just a geometry constructed in the Geant4 manner, and particles are practically transported in this geometry. On the other hand, an EGS4 world is composed of a single "virtual" volume with no dimensions (huge enough) and a medium dynamically updated at every step. This volume is used only for actual reaction of EGS4 processes. The practical work flow at tracking time is as follows:

1. A Geant4 material of a current volume is converted to a corresponding EGS4 medium through the *g4-egs4* material map.
2. Medium information of the EGS4 volume is updated.
3. Kinematical information in the EGS4 stack is updated from a Geant4 track.
4. An EGS4 process is invoked and recorded to the EGS4 stack.
5. The stack information of EGS4 is evacuated to a change of particle in the Geant4 representation.

A schematic view of the flow is also shown in Figure 1.

### 3.4. User Interface

What is provided from users' side of view? At first, we provide another user physics list for using EGS4 processes. To use EGS4 processes, any existing Geant4 geometry codes don't have to be modified. As for materials, users have to prepare EGS4 materials by themselves. Practically, users define PEGS materials in a PEGS input file and execute PEGS to generate a cross section file. In addition, uses have to give assignments for which PEGS material corresponds to a Geant4 material used in users geometry.

All operations for setting user parameters and flags controlling behavior of EGS4 programs can be performed through Geant4 command line interface. So, users don't have to recompile any programs to modify user parameters and flags.

### 4. CURRENT STATUS AND PLAN

The development was carried out under Linux system (SuSE Linux 8.1). We used gcc and g77 of version 3.2 and Geant4 of version 5.0 with patch01. The version of EGS4 used was a version of distributed at the KEK site[2].

The first implementation was finished, and we have checked basic functionalities. In Figure 2, a sample event display of electromagnetic shower event is shown. As for robustness of the system, several million events in a simple geometry were successfully generated without any problems. Detail system check and benchmark tests are started as the next target. We have plans for comparisons between EGS itself and the EGS4-G4 interface and between G4 and EGS4 with this interface on common

---
[2] ftp://ftp.kek.jp/kek/kek_egs4/egs4unix_kek/

geometries. And also, technical study for including PRESTA is planned for more precise simulation.

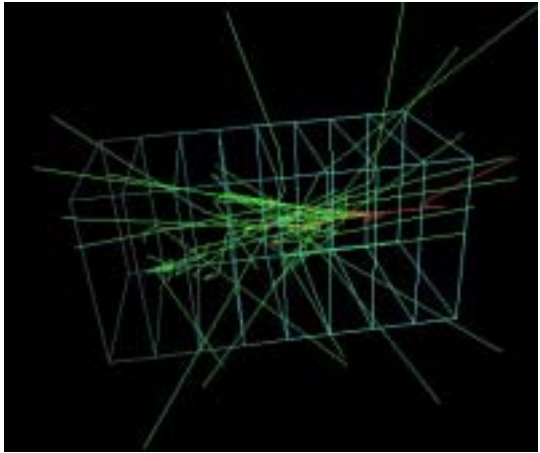

Figure 2: Sample event display. Five electrons with $E_k$=50MeV into a slab geometry composed of several materials. Cutoff energy is set to $E_k$=10keV.

In summary, we have successfully developed the first version of an interface between Geant4 and EGS4 based on object-oriented approach. This work is a proof of flexibility and expansibility of the Geant4 framework. By means of this interface, EGS4 users will be able to share Geant4 powerful resources, such as geometry description, tracking, etc.

**THMT006**